\documentclass[aps,prl,twocolumn,reprint,groupedaddress,superscriptaddress]{revtex4-1}

\usepackage{graphicx}
\usepackage{amsmath}

\begin{document}

%\preprint{}

\title{High-precision measurement of microwave electric field by cavity-enhanced critical behavior in a many-body Rydberg atomic system}


\affiliation{State Key Laboratory of Quantum Optics Technologies and Devices, and
Institute of Opto-Electronics, Shanxi University, Taiyuan 030006,
China}
\affiliation{Institute for history of science and technology, Shanxi University, Taiyuan 030006, China}
\affiliation{Collaborative Innovation Center of Extreme Optics, Shanxi University,
Taiyuan 030006, China}

\author{Qinxia Wang}
\thanks{These authors contributed equally to this work.}
\affiliation{State Key Laboratory of Quantum Optics Technologies and Devices, and
Institute of Opto-Electronics, Shanxi University, Taiyuan 030006,
China}
\affiliation{Institute for history of science and technology, Shanxi University, Taiyuan 030006, China}

\author{Yukang Liang}
\thanks{These authors contributed equally to this work.}
\affiliation{State Key Laboratory of Quantum Optics Technologies and Devices, and
Institute of Opto-Electronics, Shanxi University, Taiyuan 030006,
China}

\author{Zhihui Wang}
\thanks{These authors contributed equally to this work.}
\affiliation{State Key Laboratory of Quantum Optics Technologies and Devices, and
Institute of Opto-Electronics, Shanxi University, Taiyuan 030006,
China}
\affiliation{Collaborative Innovation Center of Extreme Optics, Shanxi University,
Taiyuan 030006, China}

\author{Shijun Guan}
\affiliation{State Key Laboratory of Quantum Optics Technologies and Devices, and
Institute of Opto-Electronics, Shanxi University, Taiyuan 030006,
China}

\author{Pengfei~Yang}
\affiliation{State Key Laboratory of Quantum Optics Technologies and Devices, and
Institute of Opto-Electronics, Shanxi University, Taiyuan 030006,
China}
\affiliation{Collaborative Innovation Center of Extreme Optics, Shanxi University,
Taiyuan 030006, China}

\author{Pengfei Zhang}
\affiliation{State Key Laboratory of Quantum Optics Technologies and Devices, and
Institute of Opto-Electronics, Shanxi University, Taiyuan 030006,
China}
\affiliation{Collaborative Innovation Center of Extreme Optics, Shanxi University,
Taiyuan 030006, China}

\author{Gang Li}
\email{gangli@sxu.edu.cn}
\affiliation{State Key Laboratory of Quantum Optics Technologies and Devices, and
Institute of Opto-Electronics, Shanxi University, Taiyuan 030006,
China}
\affiliation{Collaborative Innovation Center of Extreme Optics, Shanxi University,
Taiyuan 030006, China}

\author{Tiancai Zhang}
\affiliation{State Key Laboratory of Quantum Optics Technologies and Devices, and
Institute of Opto-Electronics, Shanxi University, Taiyuan 030006,
China}
\affiliation{Collaborative Innovation Center of Extreme Optics, Shanxi University,
Taiyuan 030006, China}


\begin{abstract}
It has been demonstrated that the Rydberg criticality in a many-body atomic system can enhance the measurement sensitivity of the microwave electric field by increasing the Fisher information. In our previous work, we proposed and experimentally verified that the Fisher information near the critical point can be increased by more than two orders of magnitude with the Rydberg atoms coupled with an optical cavity compared with that in free space. Here we demonstrate the precision measurement of the microwave electric field by cavity-enhanced critical behavior. We show that the equivalent measurement sensitivity of the microwave electric field can be enhanced by an order of magnitude compared with that in free space. The obtained sensitivity can be enhanced to 2.6 nV/cm/Hz$^{1/2}$.
\end{abstract}


%\keywords{}

\maketitle

Rydberg atoms \cite{ref1, ref2} are susceptible to external electric fields from DC to optical frequencies \cite{ref3, ref4} due to their large electric dipole moments and abundant transitions. Therefore, they can be used naturally to perform international system of units (SI)-traceable and highly sensitive microwave (MW) electric field measurements \cite{ref5,ref6,ref7,ref8,ref9,ref10,ref11,ref12}. The available sensitivity with a hot atomic vapor is mainly limited by the Rydberg atomic spectral signal-to-noise ratio (SNR). Several methods, such as superheterodyne \cite{ref6}, Mach-Zehnder interferometer \cite{ref8}, amplitude modulation, choosing high Rydberg states \cite{ref10}, etc., have been adopted to enhance the SNR and the sensitivity can arrive at 5.1 nV/cm/Hz$^{1/2}$. 
Recently, the dynamics of a non-equilibrium phase transition (NEPT) \cite{ref13,ref14,ref15} associated with the Rydberg atom have been introduced to measure the MW electric field \cite{ref7}. The NEPT refers to the switching between two non-equilibrium states of a Rydberg atom system driven by laser field. The transition is induced by the competition among the long-range interactions between Rydberg atoms \cite{ref14,ref16,ref17}, the driving strength of the laser field, and the dissipation of the Rydberg state.
Due to the Rydberg interaction, the system can reside in either state under the same driving condition and gives two distinct output light fields. 
The phenomenon is generally termed optical bistability (OB) \cite{ref7,ref12,ref13,ref16,ref20,ref18,ref19}. 
When the frequency or intensity of the driving field is scanned, the system can jump from one state to the other.
The jumping direction depends on the scan direction, and a hysteresis loop is usually observed. The speed of the jump is very fast in a hot atom vapor cell because of the fast dynamics of the light-atom and atom-atom interactions. Therefore, the jump exhibits a sharp edge in the spectrum when the light frequency is scanned. 
A critical point of the NEPT can be defined as the driving light frequency at the jump point. The sharp edge can be adopted as a fine frequency discriminator to enhance the measurement of the MW field \cite{ref21,ref22,ref23,ref24}. The demonstrated sensitivity of MW measurement using the NEPT is 49 nV/cm/Hz$^{1/2}$.

In principle, a stronger interaction results in a sharper edge of the transition. The optical cavity can enhance the light-matter interaction \cite{ref25,ref26} and induce long-range interaction between atoms \cite{ref27,ref28,ref29,ref30,ref31}. The cavity-enhanced Rydberg OB has been experimentally demonstrated with a thermal atom vapor cell \cite{ref32}. The slope of the phase transition near the critical point and the classical Fisher information (FI) \cite{ref33} for the frequency discrimination can be increased more than 10 and 100 times compared to those without the cavity. In this article, we report the measurement of the MW electric field using cavity-enhanced NEPT with the method introduced in Ding's work \cite{ref7}. An order of magnitude improvement on the measurement sensitivity is demonstrated in comparison to that with atom vapor in free space. 
The achieved highest sensitivity is 2.6 nV/cm/Hz$^{1/2}$.

\section{The experimental scheme}\label{sec:2}

\begin{figure}
\centering
\includegraphics[scale=0.5]{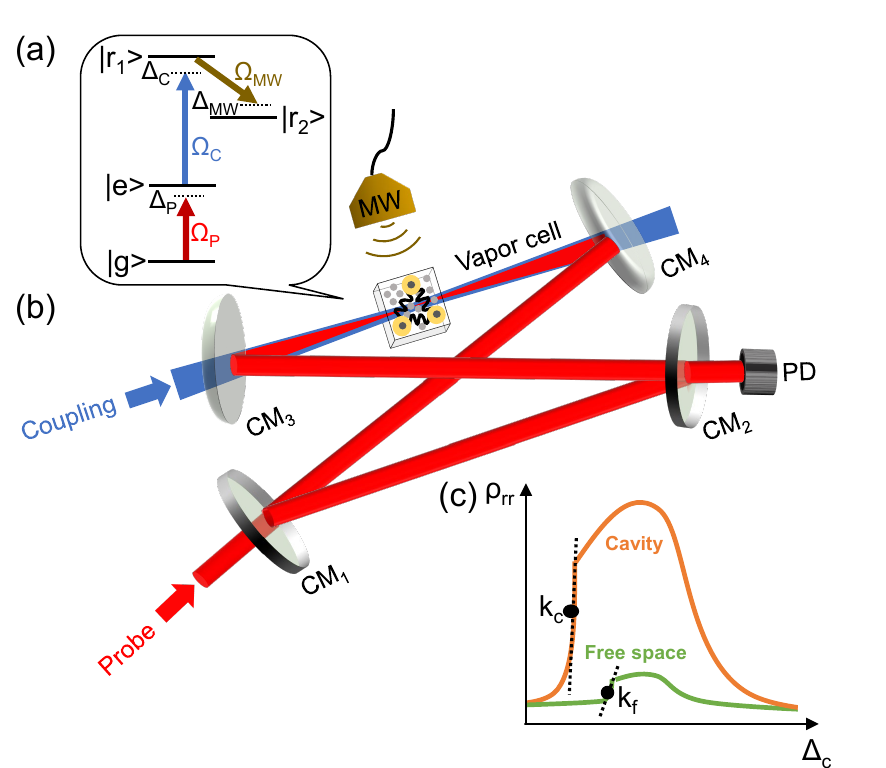}
\caption{Experimental sketch of microwave electrometry based on cavity-enhanced NEPT with Rydberg atom vapor. (a) Energy level scheme of the Cs atom used in the experiment with $|g\rangle \equiv 6{S_{1/2}}|F = 4\rangle \to |e\rangle \equiv 6{P_{3/2}}|F' = 5\rangle$, $|e\rangle\to |r_1\rangle \equiv 47{D_{5/2}}\rangle$, and $|r_2\rangle \equiv 48{P_{3/2}}\rangle$. (b) Experimental setup. CM: cavity mirror; PD: photodiode; MW: microwave antenna. (c) Comparison of NEPT with Rydberg atom in a cavity and free space. The cavity-enhanced NEPT exhibits steeper slopes ($k_{c}$) compared to the free-space case ($k_{f}$).}
\label{fig:example1}
\end{figure}

The experimental scheme of microwave electrometry based on a cavity-enhanced NEPT with Rydberg atom vapor is shown in Fig. \ref{fig:example1}. Figure \ref{fig:example1}(a) is the energy level scheme of the cesium (Cs) atom adopted in the experiment. The four atomic states are the ground state $|g\rangle \equiv 6{S_{1/2}}|F = 4\rangle$, excited state $|e\rangle \equiv 6{P_{3/2}}|F' = 5\rangle$, and two Rydberg states $|r_1\rangle \equiv 47{D_{5/2}}\rangle$, $|r_2\rangle \equiv 48{P_{3/2}}\rangle$. A probe (coupling) laser couples atomic transition $|g\rangle\to |e\rangle$ ($|e\rangle\to |r_1\rangle$) with Rabi frequency $\Omega _{p}$ ($\Omega _{c}$) and frequency detuning $\Delta_{p}$ ($\Delta_{c}$). A MW field couples Rydberg transition $|r_1\rangle\to |r_2\rangle$ with Rabi frequency $\Omega_{MW}$ and frequency detuning $\Delta_{MW}$. We obtained the OB transmission spectra by scanning the frequency of the coupling laser with the frequency of the probe laser fixed. The strength of external MW fields can be measured by determining the power of the transmitted probe laser at the edge of the NEPT. 

The experimental setup is illustrated in Fig. \ref{fig:example1}(b). 
An optical cavity with a bow-tie configuration is designed to enhance the intensity of the probe laser. 
The finesse of the empty cavity is approximately 85, which is reduced to approximately 20 when the atomic vapor cell with a size of 15 mm is taken into account, owing to the atom absorption and the reflection of the cell walls.
The probe laser is injected into the optical cavity with a bow-tie configuration via cavity mirror 1 (CM1). 
The cavity length is actively stabilized and the frequency is resonant with the probe laser. The coupling laser does not circulate inside the cavity due to the high transmission of the cavity mirrors. It transits the cavity and the atomic vapor via cavity mirrors 3 and 4 (CM3 and CM4). 
The probe and coupling lasers are counter-propagating through the atomic vapor to reduce the Doppler effect \cite{ref34} on the Rydberg spectra. 
The beam waists of the probe and coupling lasers are located between CM3 and CM4 with sizes of 50 and 80 $\mu$m, respectively. 
The leakage of the probe laser from cavity mirror 2 (CM2), which indicates the population of the Rydberg state $\rho _{rr}$, is received by a high-bandwidth photodetector PD2 (C30659-900-R8A, Perkin Elmer). 
The transmissive spectra of the system are obtained by recording the leakage of the probe light while the frequency of the coupling laser ($\Delta_c$) is scanned. 
A microwave antenna (LB-SJ-60180-SF, A-INFOMW) is placed 20 cm above the vapor cell, and the microwave electric field calibration at higher microwave power was performed by using the microwave-induced Autler-Townes (AT) splitting \cite{ref5}.

\section{ Measured derivative and FI}\label{sec:3}

In our experiments, the transmissive spectra with and without the cavity are recorded with the same experimental setup by switching an optical shutter \cite{ref32}, which is placed between CM1 and CM2. Fig. \ref{fig:example1}(c) shows the typical curves of the spectra with high probe light power. 
The NEPT with steep transition edge can be clearly observed. The black dashed lines are drawn using the maximum slopes of the spectrum near the critical point, and the slopes are also indicated by the symbols $k_{c}$ (cavity) and $k_{f}$ (free space).  
As we have proved in our previous work \cite{ref32}, a cavity-enhanced system has a steeper slope than that in the free space. Therefore, the cavity-enhanced system has a larger optical response to a small frequency shift at the phase transition edge, which implies a higher sensitivity for microwave electric field measurement. 

The precision of measuring the frequency shift $\Delta$ is described by the Cramér-Rao bound, $\delta\Delta \geq1/{\sqrt{\text{$F$}(\Delta)}}$, where $\text{$F$}(\Delta)$ is the Fisher information (FI) of the measured intensity. 
The larger FI indicates higher measurement precision. Therefore, FI can be used to quantify the sensitivity in precision microwave electric field measurements \cite{ref7}.
FI is calculated by $\text{$F$}(\Delta) = k^{2} /\text{Var}(\mu)$, 
where $k$ represents the maximum slope of the spectrum around the critical point, and Var($\mu$) denotes the noise variance.
Consequently, by analyzing the slope of the spectrum within a narrow time interval around the critical point, the sensitivity can be explicitly expressed in terms of the FI.

\begin{figure}
\centering
\includegraphics[scale=1.2]{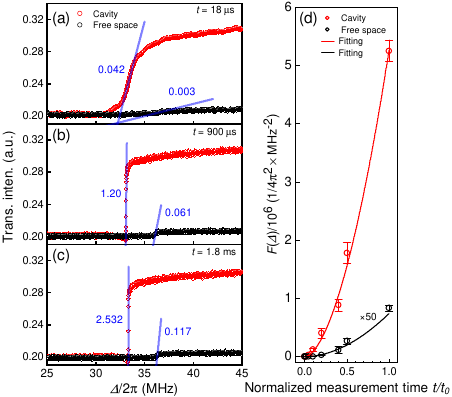}
\caption{(Color online) Transmissive spectra and associated Fisher information. (a)-(c) show the transmissive intensity versus the coupling laser detuning in the free space (black lines) and the cavity (red lines) with the total data acquisition time of 18 $\mu$s (a), 900 $\mu$s (b) and 1.8 ms (c). The blue lines and numbers show the maximum slopes $k$ near critical points. (d) shows the FI for different normalized measurement times of $t/t_{0}$, $t_{0}=1.8$ ms. The FI for the free space case is displayed with a magnification factor of 50. The red and black curves are the data fittings by $F = A(t/t_{0} )^{\lambda}$ with $A = 5.27\times10^{6}$ MHz$^{-2}$ and $\lambda = 1.76$ for cavity, $A = 8.02\times 10^{3}$ MHz$^{-2}$ and $\lambda = 1.73$ for free space.} 
\label{fig:example2}
\end{figure}

The FI also depends on the integration time $t$ of the measurement system \cite{ref22}, defined as the time that the coupling laser explores a small frequency interval around each detuning. Due to the critical slowing down near phase transitions, FI has a non-integer power \cite{ref7} dependence on $t$. Thus, we measured the FI of the free space and the cavity at different integration times $t$ as shown in Fig. \ref{fig:example2}. 
Figure \ref{fig:example2}(a)-(c) show the transmissive intensity versus the coupling laser detuning in the free space and the cavity. It is obvious that our system displays a second-order dynamical phase transition between two stationary states with different excitation densities \cite{ref15}. We extracted the maximum slope value $k$ for each spectrum, which is marked in each subfigure. The Var($\mu$) associated with each curve is also measured. 
The transmission slope near the critical point becomes steeper as the integration time $t$ increases. The slope with the cavity is steeper than that in free space. The maximum slope $k$ is enhanced approximately 21 times with the cavity compared to that in free space, whereas the noise is only enlarged by 1.2 times on average. The additional noise mainly comes from the instability of the cavity, which depends on the performance of the cavity-locking loops. 

The extracted values of FI at different integration times are shown in Fig. \ref{fig:example2}(d). 
The FI depends on the integrating time non-linearly, and the data fit well by $F=A(t/t_{0})^{\lambda}$ with $t_{0}=1.8$ ms.  
Comparing the data with the cavity to that in the free space, we can see that an enhancement ratio of up to 400 can be achieved at an integrating time of 1.8 ms.
Thus, more information can be extracted using a cavity-enhanced system than a free-space system. 
From the trend of the data, more information can be extracted by using a longer integration time.

\section{NEPT signals under electric fields with different amplitudes}\label{sec:4}

\begin{figure}
\centering
\includegraphics[scale=0.3]{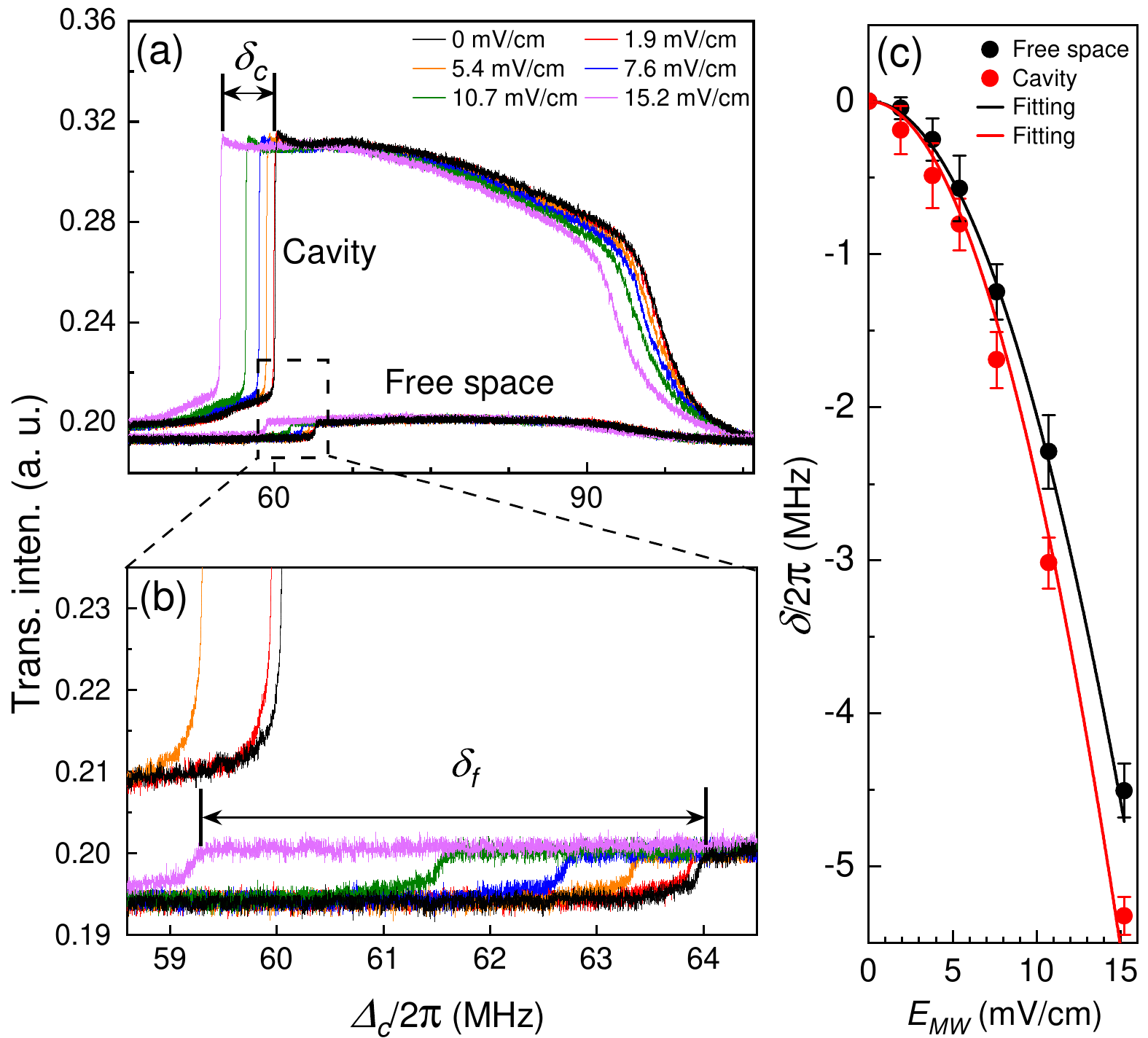}
\caption{(Color online) Measurement of the shifted critical point due to the strength of the MW field. (a) Rydberg OB spectra in the free space and cavity with the strength of the MW field scanned. The frequency is set as $2\pi\times6.76$ GHz and the amplitude of the electric field is scanned $E_{MW} = 0$ mV/cm to 15.2 mV/cm with probe Rabi frequency $\Omega _{p}/2\pi = 19$ MHz and coupling Rabi frequency $\Omega _{c}/2\pi = 20$ MHz. The direction of scanning $\Delta_{c}$ is from red-detuning towards the blue-detuning and the sweep rate is $2\pi\times0.01$ MHz$/\mu$s. (b) The enlarged view near the critical point of the free space case. (c) The shift of the critical point in the cavity (red dots) and free space (black dots) as a function of E$_{MW}$. The data are fit by $\delta = -(\Delta_{MW}/2+\sqrt{\Delta_{MW}^{2}+\Omega_{MW}^{2}}/2)-\eta\Omega_{MW}^{2}$ with experimental parameters of $\Delta_{MW} = -2\pi\times 200$ MHz. Here, $\Omega_{MW}$ is the Rabi frequency of the MW electric field and it can be expressed as $\Omega_{MW} = \mu_{0}E_{MW}/$\( \hbar\), where $\mu_{0 } =$ 2938\( e \)\( a_0 \) is the dipole momentum, \( e \) is the electronic charge and \( a_0 \) is the Bohr radius. The fitted coefficient are $\eta_{cavity} = 5.25\times10^{-4}$ MHz$^{-1}$ (red) and $\eta_{free-space} = 2.23\times10^{-4}$ MHz$^{-1}$(black).} 
\label{fig:example3}
\end{figure}

To further apply this cavity-enhanced critical response to sense the MW, a MW field is applied to the atomic vapor to drive the atomic transition $47{D_{5/2}}\to 48{P_{3/2}}$ with frequency detuning $\Delta _{MW}/2\pi = -200$ MHz.
The responses of the NEPT on the MW with a series of electric fields are shown in Fig. \ref{fig:example3} (a) and (b), where the signals with and without the cavity are displayed for comparison.
The signals without the MW are drawn with black lines. 
The application of the MW will induce an a.c. Stark shift \cite{ref36} on the atomic transition, which will be finally reflected on the NEPT signal as a small frequency shift.
The direction of the frequency shift is related to the polarizability of the Rydberg atom and the detuning of the MW electric field \cite{ref7}.
The amount of the frequency shift can be seen as a ruler to measure the applied MW.
The edge of the NEPT signal is the ruler tick. 
Ding et al. have demonstrated that the edge of NEPT in free space can provide fine sticks to measure the MW \cite{ref7}.
Here we see that the edge of cavity-enhanced NEPT can provide finer sticks than that in free space.
Therefore, the cavity-enhanced system has the potential to measure the MW with higher sensitivity than the NEPT-based MW sensor in free space \cite{ref7}. 

The relationship between the frequency shift $\delta$ of the NEPT edge and the electric field strength $E_{MW}$ of the applied MW is then studied, and the data are summarized in Fig. \ref{fig:example3}(c), where the black and red dots represent the data in free space and cavity, respectively. 
The frequency shift of the NEPT signal in the cavity is greater than that in the free space under the same MW field, which means a bigger Rydberg interaction in the cavity \cite{ref32}.

Both sets of data can be fitted by the formula \cite{ref7} $\delta = -\left(\Delta_{MW}/2+\sqrt{\Delta_{MW}^{2}+\Omega_{MW}^{2}}/2\right)-\eta\Omega_{MW}^{2}$, which describes the dependence of frequency shifts $\delta$ of the transmission spectra on the Rabi microwave frequency $\Omega_{MW}$. The first term on the right-hand side of the formula represents the frequency shift of the Rydberg energy levels due to the a.c. Stark effect, which is the same in both free space and the cavity. The second term accounts for the frequency shift caused by interatomic interactions between Rydberg atoms \cite{ref13}. Due to the stronger atomic interactions within the cavity, the resulting frequency shift is larger than that observed in free space.
The coefficient $\eta$ gives the strength of the Rydberg interaction.
From the data fitting, we get the $\eta_\text{c} = 5.25\times10^{-4}$ MHz$^{-1}$ and $\eta_\text{f} = 2.23\times10^{-4}$ MHz$^{-1}$ for the cases in the cavity and free space, respectively.
Knowing the relation, the electrical field strength of the applied MW can be measured by reading the frequency shift of the NEPT signal.

\section{Measured the sensitivity of the MW electric field }\label{sec:5}

The measurement sensitivity of the MW electric field is a key factor in evaluating the electrometer's performance. For the NEPT-based electrometer, the sensitivity is determined by the slope of the phase transition edge. Our cavity-enhanced NEPT system can further enhance the sensitivity. To estimate the sensitivity, we follow the methods used in Ding's work \cite{ref7}.

\begin{figure}
\centering
\includegraphics[scale=0.3]{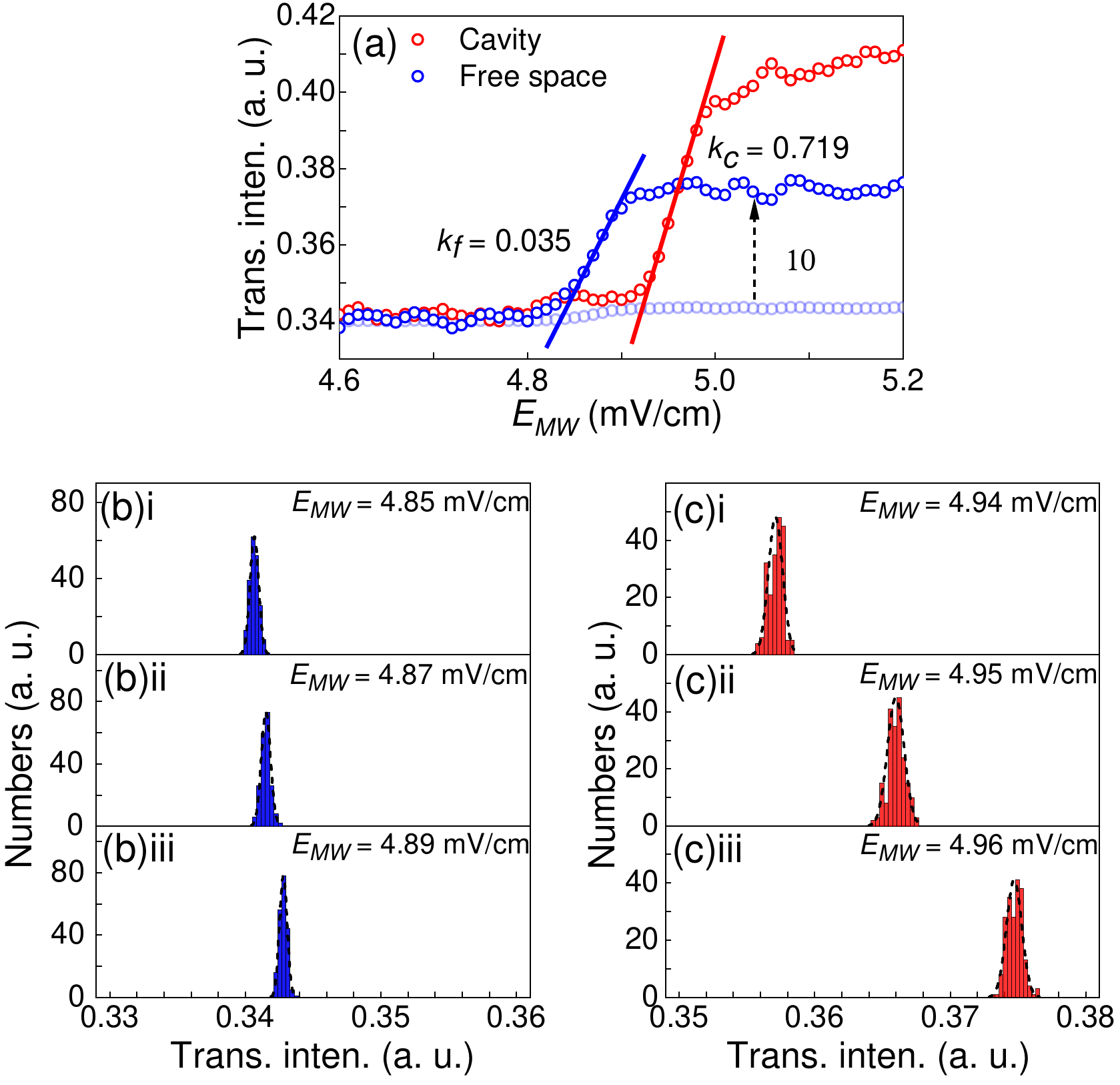}
\caption{(Color online)Transmission of the Rydberg system under different amplitudes of MW electric field. (a) The transmissive intensity of the probe laser in the free space (blue circle) and cavity (red circle). The amplitude $E_{MW}$ is scanned from 4.6 mV/cm to 5.2 mV/cm in steps of 10 $\mu$V/cm per 5 $\mu$s. The light blue line is the original signal in free space and the blue line is the amplified curve for a clear view. The blue and red solid lines are and the symbols $k_{f}$ and $k_{c}$ indicate the maximum slopes. The sub-figures (b)i - (b)iii and (c)i - (c)iii display the histograms of the transmissive intensity around the critical point in the free space and cavity with different electric strengths of the MW field. The dashed lines are fitted curves by Gaussian function.}
\label{fig:example4}
\end{figure}

We measured the NEPT signals with the amplitude of the MW electric field scanned while the frequencies of the probe and coupling lasers were fixed near the critical point. The results are shown in Fig. \ref{fig:example4}(a), in which the NEPT signal without the cavity is also shown for comparison. The amplitude $E_{MW}$ of the MW electric field is scanned from 4.6 mV/cm to 5.2 mV/cm with a step of 10 $\mu$V/cm per 5 $\mu$s. We can see that the NEPT signals are sensitive to the electrical field of MW around $E_{MW} = $ 4.87 mV/cm and $E_{MW} = $ 4.95 mV/cm for the free-space and cavity cases, respectively. This position can be tuned by the frequency detuning of the coupling light \cite{ref7}. To evaluate the sensitivity of MW electrical field measurement, we extract the maximum slopes at the half edge of the NEPT signal. The values are $k_\text{c} = 0.719$ and $k_\text{f} = 0.035$ for the cavity and free-space cases. 
By using the cavity, $k$ is enhanced approximately 21 times compared to the signal in free space, and the noise variance is enlarged approximately 1.1 times. 
The sensitivity can be obtained by analyzing the histogram of the NEPT signal at different MW amplitudes along the edge of the phase transition.
Figure \ref{fig:example4}(b) and (c) display the results for the cases in free space and cavity, respectively, where the acquisition time is 5 $\mu$s for every data point. 
The widths of the histogram peaks are $\delta E_{MW} = 22.5$ $\mu$V/cm for the free-space case and $\delta E_{MW} = 1.19$ $\mu$V/cm for the cavity case. 
The corresponding sensitivities are 50.3 nV/cm/Hz$^{1/2}$ in free space and 2.6 nV/cm/Hz$^{1/2}$ in cavity.
The sensitivity is improved by an order of magnitude with the optical cavity. 
It could be further improved by optimizing the fineness and locking loop of the cavity \cite{ref32}.

\section{Conclusion}\label{sec:6}

In summary, we have experimentally demonstrated high-precision MW electric field measurements using a cavity-enhanced NEPT in a cesium Rydberg atom vapor cell, achieving a significant improvement in the sensitivity. By adopting an optical cavity, we enhanced the interatomic Rydberg interactions, resulting in a larger signal and steeper jumping (phase transition) slopes in the OB signal. 
The classical FI was boosted by a factor of 400 compared to the free-space case.
Compared to the measurement sensitivity of 50.3 nV/cm/Hz$^{1/2}$ with NEPT in free space, the sensitivity was enhanced by an order of magnitude, reaching 2.6 nV/cm/Hz$^{1/2}$. 
To further enhance sensitivity, future efforts can focus on optimizing cavity design and implementing noise suppression techniques. 
Our results show the great potential of cavity-enhanced Rydberg atomic systems for ultrasensitive MW field detection and contribute to the advancement of quantum metrology. 
The approach holds great promise for applications in precision sensing, quantum information processing, and the study of critical phenomena in many-body systems. 

\begin{acknowledgements}
This work was supported by National Key Research and Development Program of China (2021YFA1402002); National Natural Science Foundation of China (U21A6006, U21A20433, 92465201, 12104277, 12104278, 12474360 and 92265108); Fund for Shanxi “1331 Project” Key Subjects; Postdoctoral Fellowship Program of CPSF (GZC20240960).
\end{acknowledgements}


\begin{thebibliography}{0}%
\makeatletter
\providecommand \@ifxundefined [1]{%
 \@ifx{#1\undefined}
}%
\providecommand \@ifnum [1]{%
 \ifnum #1\expandafter \@firstoftwo
 \else \expandafter \@secondoftwo
 \fi
}%
\providecommand \@ifx [1]{%
 \ifx #1\expandafter \@firstoftwo
 \else \expandafter \@secondoftwo
 \fi
}%
\providecommand \natexlab [1]{#1}%
\providecommand \enquote  [1]{``#1''}%
\providecommand \bibnamefont  [1]{#1}%
\providecommand \bibfnamefont [1]{#1}%
\providecommand \citenamefont [1]{#1}%
\providecommand \href@noop [0]{\@secondoftwo}%
\providecommand \href [0]{\begingroup \@sanitize@url \@href}%
\providecommand \@href[1]{\@@startlink{#1}\@@href}%
\providecommand \@@href[1]{\endgroup#1\@@endlink}%
\providecommand \@sanitize@url [0]{\catcode `\\12\catcode `\$12\catcode `\&12\catcode `\#12\catcode `\^12\catcode `\_12\catcode `\%12\relax}%
\providecommand \@@startlink[1]{}%
\providecommand \@@endlink[0]{}%
\providecommand \url  [0]{\begingroup\@sanitize@url \@url }%
\providecommand \@url [1]{\endgroup\@href {#1}{\urlprefix }}%
\providecommand \urlprefix  [0]{URL }%
\providecommand \Eprint [0]{\href }%
\providecommand \doibase [0]{http://dx.doi.org/}%
\providecommand \selectlanguage [0]{\@gobble}%
\providecommand \bibinfo  [0]{\@secondoftwo}%
\providecommand \bibfield  [0]{\@secondoftwo}%
\providecommand \translation [1]{[#1]}%
\providecommand \BibitemOpen [0]{}%
\providecommand \bibitemStop [0]{}%
\providecommand \bibitemNoStop [0]{.\EOS\space}%
\providecommand \EOS [0]{\spacefactor3000\relax}%
\providecommand \BibitemShut  [1]{\csname bibitem#1\endcsname}%
\let\auto@bib@innerbib\@empty
%</preamble>
\end{thebibliography}%


\begin{thebibliography}{99}

\bibitem {ref1}C. S. Adams, J. D. Pritchard, and J. P. Shaffer, J. Phys. B: At. Mol. Opt. Phys. 53, 012002 (2019).

\bibitem {ref2}M. Saffman, T. G. Walker, and K. Mølmer, Rev. Mod. Phys. 82, 2313–2363 (2010).

\bibitem {ref3}A. Facon, E. K. Dietsche, D. Grosso, S. Haroche, J. M. Raimond, M. Brune, and S. Gleyzes, Nature 535, 262–265 (2016).

\bibitem {ref4}K. C. Cox, D. H. Meyer, F. K. Fatemi, and P. D. Kunz, Phys. Rev. Lett. 121, 110502 (2018).

\bibitem {ref5}J. A. Sedlacek, A. Schwettmann, H. Kübler, R. Löw, T. Pfau, and J. P.
Shaffer, Nat. Phys. 8, 819–824 (2012).

\bibitem {ref6}M. Y. Jing, Y. Hu, J. Ma, H. Zhang, L. J. Zhang, L. T. Xiao, and S. T.
Jia, Nat. Phys. 16, 911–915 (2020).

\bibitem {ref7}D. S. Ding, Z. K. Liu, B. S. Shi, G. C. Guo, K. Mølmer, and C. S. Adams, Nat. Phys. 18, 1447 (2022).

\bibitem {ref8}W. G. Yang, M. Y. Jing, H. Zhang, L. J. Zhang, L. T.Xiao, and S. T. Jia, Phys. Rev. Appl. 19, 064021 (2023).

\bibitem {ref9}H. T. Tu, K. Y. Liao, G.-D. He, Y. F. Zhu, S. Y. Qiu, H. Jiang, W. Huang, W. Bian, H. Yan, and S. L. Zhu, arXiv:2307.15617 (2023).

\bibitem {ref10}M. H. Cai, S. H. You, S. S. Zhang, Z. S. Xu, and H. P. Liu, Appl. Phys. Lett. 122, 161103 (2023).

\bibitem {ref11}K. Dixon, K. Nickerson, D. W. Booth, and J. P. Shaffer, Phys. Rev. Appl. 19, 034078 (2023).

\bibitem {ref12}K. D. Wu, C. W.  Xie, C. F. Li, G. C. Guo, C.L. Zou, and G.Y. Xiang, Sci. Adv. 10, eado8130 (2024).

\bibitem {ref13}D. S. Ding, H. Busche, B. S. Shi, G. C. Guo, and C. S. Adams, Phys. Rev. X 10, 021023 (2020).

\bibitem {ref14}C. Carr, R. Ritter, C. G. Wade, C. S. Adams, and K. J. Weatherill, Phys. Rev. Lett. 111, 113901 (2013).

\bibitem {ref15}M. Marcuzzi, E. Levi, S. Diehl, J. P. Garrahan, and I. Lesanovsky, Phys. Rev. Lett. 113, 210401 (2014).

\bibitem {ref16}N. R. de Melo, C. G. Wade, N. Šibali´c, J. M. Kondo, C. S. Adams, and
K. J. Weatherill, Phys. Rev. A 93, 063863 (2016).

\bibitem {ref17}T. Baluktsian, B. Huber, R. Löw, and T. Pfau, Phys. Rev. Lett.
110, 123001 (2013).

\bibitem {ref18}T. E. Lee, H. Häffner, and M. C. Cross, Phys. Rev. Lett. 108, 023602 (2012).

\bibitem {ref19}D. Weller, A. Urvoy, A. Rico, R. Löw, and H. Kübler, Phys. Rev. A 94, 063820 (2016).

\bibitem {ref20}J. He, X. Wang, X. Wen, and J. M. Wang, Opt. Express 28, 33682–33689 (2020).

\bibitem {ref21}L. Garbe, M. Bina, A. Keller, M. G. A. Paris, and S. Felicetti, Phys. Rev. Lett. 124, 120504 (2020).

\bibitem {ref22}T. Ilias, D. Yang, S. F. Huelga, and M. B. Plenio, PRX Quantum 3, 010354
(2022).

\bibitem {ref23}Y. M. Chu, S. L. Zhang, B. Y. Yu, and J. M. Cai, Phys. Rev. Lett. 126, 010502 (2021).

\bibitem {ref24}V. Montenegro, U. Mishra, and A. Bayat, Phys. Rev. Lett. 126,
200501 (2021).

\bibitem {ref25}H. J. Kimble, Physica Scripta 1998, 127 (1998).

\bibitem {ref26}S. M. Dutra, Cavity Quantum Electrodynamics (John Wiley and Sons, Ltd,
2004).

\bibitem {ref27}R. Mottl, F. Brennecke, K. Baumann, R. Landig, T. Donner, and T. Esslinger, Science 336, 1570–1573 (2012).

\bibitem {ref28}S. Schütz and G. Morigi, Phys. Rev. Lett. 113, 203002 (2014).

\bibitem {ref29}E. J. Davis, G. Bentsen, L. Homeier, T. Li, and M. H. Schleier-Smith, Phys. Rev. Lett. 122, 010405 (2019).

\bibitem {ref30}J. Muniz, D. Barberena, R. J. Lewis-Swan, D. J. Young, J. R. K. Cline, A. M. Rey, and J. K. Thompson, Nature 580, 602 (2020).

\bibitem {ref31}A. Periwal, E. S. Cooper, P. Kunkel, J. F. Wienand, E. J. Davis, and M. Schleier-Smith, Nature 600, 630 (2021).

\bibitem {ref32}Q. X. Wang, Z. H. Wang, Y. X. Liu, S. J. Guan, J. He, C. L. Zou, P. F. Zhang, G. Li, and T. C. Zhang,  Opt. Lett. 48, 2865–2868 (2023).

\bibitem {ref33}L. Pezzè, A. Smerzi, M. K. Oberthaler, R. Schmied, and P. Treutlein, Rev. Mod. Phys. 90, 035005 (2018).

\bibitem {ref34}B. D. Yang, J. Gao, T. C. Zhang, and J. M. Wang, Phys. Rev. A 83, 013818 (2011).

%由于{ref35和ref15重复}，已将{ref35}删除%

\bibitem {ref36}P. Bohlouli-Zanjani, J. A. Petrus, and J. D. D. Martin, Phys. Rev. Lett. 98, 203005 (2007).

\end{thebibliography}
\end{document}